\documentclass[12pt]{article}
\usepackage[utf8]{inputenc}
\usepackage[T1]{fontenc}
\usepackage{graphicx,authblk,hyperref,url,classif2}
\pdfoutput=1

\author[1]{Alexey Shipunov}

\affil[1]{Minot State University, Minot ND, 58707}

\title{``Numerical ranks'' to improve biological nomenclature of higher groups}

\date{}

\newenvironment{Hang}{\leftskip2em\parindent-2em\parskip1ex\mbox{}}{}

\begin{document}

\maketitle

\begin{abstract}

Simple method to improve traditional approaches to name taxa of higher ranks has been proposed. Instead of base name + postfixes (which vary between codes of nomenclature and sometimes not standardized at all), use \textbf{numerical prefixes}, where numbers designate the rank of taxon, from lowest (species $= 1$) to highest (kingdom $= 7$). This ``numerical ranks'' method is not only more simple and straightforward then current practice, but also easy to extend for numerous additional uses. There is a hope that numerical ranks will facilitate creation of typified names for all higher-level taxomonic groups.

\end{abstract}

\section{Introduction}

Most codes of nomenclature (ICZN, 1999; Lapage et al., 1992; McNeil et al., 2012) regulate the construction of taxonomic names on the level of species and genus in a similar way, with binary nomenclature invented by Linnaeus (1753) is a main method of naming taxa.

Higher groups, however, are named in multiple different ways. Most frequent are traditional names (sometime with postfix, or without it), and names combined from name of genus (base name) and standardized postfix. Codes of nomenclature include detailed instructions of how to use some of these postfixes.

One of serious disadvantages of this system is non-uniformity of rules between codes. Another is producing really long names which sometimes hard to decipher in order to understand the actual rank (this is especially notorious among names of intermediate ranks, like suborders or infraclasses). In addition, base name is not always easy to recognize within this compound name, because grammar rules require alternations. Not easy also is to use intermediate categories, there taxonomists sometimes invent new postfixes which are not governed by codes.

\section{Results}

I propose here simple approach which is free from limitations mentioned above, and also easy enough to extend:

\begin{quote}
Instead of postfixes, use \textbf{numerical prefixes}, where numbers designate the rank of taxon, from lowest (species $= 1$) to highest (kingdom $= 7$).
\end{quote}

Intermediate ranks will then receive numbers with one decimal: sub- categories are .8, super- categories are .2, and infra- categories .5.

Here are some examples:

\begin{itemize}
\item[] $^7$\emph{Araneus} = Kingdom Animalia
\item[] $^3$\emph{Brassica} = Family Cruciferae
\item[] $^4$\emph{Agaricus} = Ordo Agaricales
\item[] $^{4.5}$\emph{Homo} = Infraclass Theria
\item[] $^{5.8}$\emph{Scolopendra} = Subphylum Myriapoda
\end{itemize}

Typographically, this system is implemented as a \LaTeX\ package ``classif2'' (Shipunov, 2008); package documentation contains some examples of use.

\section{Discussion}

Proposed numerical ranks system has multiple advantages:

\begin{enumerate}
\item It it logical and easy to remember
\item It does not create too long names
\item Base name is always clearly visible
\item Intermediate ranks are easy to use; if more categories needed, one can use two decimals and even more
\end{enumerate}

Since all numerical rank names require typification, we hope that from this attempt the list of typified names will finally emerge and become an essential part of higher taxa nomenclature.

Numerical ranks are easy to extend, for example, to cover contemporary ideas of the correspondence between rank and the age of taxon (e.g., Liu et al., 2017), here it is possible to create formulas which output numerical ranks in correspondence with timetrees.

Finally, similar numerical prefix approach could be used to label splits in cladistic, rank-less classifications. In all, numerical ranks make transitions between traditional and strict phylogenetic nomenclature easier.

\section{Literature cited}

\begin{Hang}

ICZN---International Commission on Zoological Nomenclature. 1999. International Code of Zoological Nomenclature. Fourth Edition. The International Trust for Zoological Nomenclature, London, UK. 306 pp.

Lapage, S.P., Sneath, P.H.A., Lessel, E.F., Skerman, V.B.D., Seeliger, H.P.R., Clark, W.A. 1992. International Code of Nomenclature of Bacteria. Bacteriological Code. 1990 Revision. American Society for Microbiology, Washington, D.C.

Linnaeus, C.V. 1753. Species plantarum. Laurentii Salvii, Holmiae.

Liu, J.K., Hyde, K.D., Jeewon, R., Phillips, A.J., Maharachchikumbura, S.S., Ryberg, M., Liu, Z.Y. and Zhao, Q. 2017. Ranking higher taxa using divergence times: a case study in Dothideomycetes. Fungal Diversity. 1--25.

McNeill, J., Barrie, F.R., Buck, W.R., Demoulin, V., Greuter, W., Hawksworth, D.L., Herendeen, P.S., Knapp, S., Marhold, K., Prado, J. and Prud’homme Van Reine, W.F. 2012. International Code of Nomenclature for algae, fungi and plants. Regnum vegetabile, 154.

Shipunov, A. 2008. \texttt{classif}2 package. CTAN, the Comprehensive TeX Archive. URL: \url{https://www.ctan.org/pkg/classif2}

\end{Hang}

\end{document}